\documentstyle[12pt]{article}

\begin{document}

\topmargin 0pt
\renewcommand{\thefootnote}{\fnsymbol{footnote}}
\newpage
\setcounter{page}{0}

\begin{titlepage}

\begin{flushright}
May 1994  \\ cond-mat/9405088
\end{flushright}

\vspace*{1.5cm}

\begin{center}
{\large \bf DEPINNING IN A RANDOM MEDIUM } \\

\vspace{1.5cm}
Harald Kinzelbach\footnote{
Electronic mail: kibach@iff078.iff.kfa-juelich.de}
and Michael L\"assig\footnote{
Electronic mail: lassig@iff011.dnet.kfa-juelich.de }  \\

\vspace{1.5cm}
{\em Institut f\"ur Festk\"orperforschung, \\
Forschungszentrum, 52425 J\"ulich, Germany}

\end{center}
\vspace{1.5cm}
\setcounter{footnote}{0}

\begin{abstract}
We develop a renormalized continuum field theory for a directed polymer
interacting with a random medium and a single extended defect.
The renormalization group is based on the operator algebra of the
pinning potential; it has novel features due to the breakdown of hyperscaling
in a random system. There is a second-order transition between a
localized and a delocalized phase of the polymer; we obtain analytic results on
its critical pinning strength and scaling exponents. Our results
are directly related to spatially inhomogeneous Kardar-Parisi-Zhang surface
growth.

\end{abstract}

\end {titlepage}

\newpage
\setcounter{footnote}{\arabic{footnote}}

Low-dimensional manifolds in media with quenched disorder are objects
encountered in a large variety of different physical realizations.
Obvious examples are interfaces in disordered bulk
media and random field systems
\cite{Interfaces.review,HuseHenley.roughening}
or magnetic flux lines in dirty
superconductors \cite{fluxlines},
but there is also a deep connection to the problem of
nonequilibrium surface growth \cite{KPZ,KrugSpohn.review}
and randomly driven hydrodynamics \cite{noisy.Burgers}.
Furthermore, the theory serves as a simple
paradigm for more complicated, fully frustrated random systems
such as spin glasses \cite{relation.spinglass}.

A one dimensional manifold in a random medium is a phenomenological
continuum model for a (single) magnetic  flux line in type-II superconductors
with impurities \cite{fluxlines,BalentsKardar.report}, where the flux line
interacts with an ensemble of quenched {\em point} defects
(represented by a random potential).
In addition to these point impurities there may also be
{\em extended} (e.g {\em columnar} or {\em planar})  defects
in the system. Experiments that systematically probe the effect of this
kind of impurities have recently become possible in high temperature
superconductors \cite{experiments.flux}. The statistics of the line
configurations is governed by an energetic  competition: point defects tend to
{\em roughen} the flux line;  it performs large transversal excursions in order
to take  advantage of locally favourable regions. An attractive extended
defect, on the other hand, suppresses these excursions  and, if it is
sufficiently strong, {\em localizes} the line to within a  finite transversal
distance $\xi_\perp$. The two regimes are separated by a  second order phase
transition where the localization length $\xi_\perp$ diverges. In contrast to
temperature-driven transitions, it involves the competition of two different
configuration energies rather than energy and entropy and is  hence governed by
a {\em zero-temperature} renormalization group fixed point.

The system is described by an effective Hamiltonian
\begin{equation}
{\cal H}  =  \int { {\rm d} t \,
\left\{  \,   \frac{1}{2}   \left(  \frac{ {\rm d} r }{{\rm d} t} \right)^2
- V (r,t)  +  \rho_0 \, \Phi (t)  \right \} } \; .
\label{H}
\end{equation}
Here $r(t)$ denotes the displacement vector of the flux line
(also called  {\em directed polymer}) in $d'$ transversal dimensions, as a
function of the longitudinal ``timelike'' coordinate $t$.
The random potential  $V (r,t)$, Gauss-distributed with
$ \overline{ V (r,t) } = 0 $ and
$ \overline{ V (r,t) V (r',t') } =
 2 \sigma^2 \, \delta^{d'} (r-r')  \delta (t-t')$, models the
quenched point disorder. Averages over disorder are denoted by
an overbar, thermal averages by brackets $ \langle \dots \rangle $. The last
term in the Hamiltonian describes the interaction with the extended defect. In
this letter, we concentrate on columnar defects,
 $\Phi (t) = \delta^{d'} \bigl( r(t) \bigr)$,
but most of the results can straightforwardly be generalized to planar defects.

This model gains additional interest since it is related to
nonequilibrium critical phenomena \cite{inhomogeneous.KPZ}. If $Z(r,t)$ denotes
the restricted partition sum over all paths ending at a  fixed given
point $(r,t)$, the ``height field''  $h(r,t) = \beta^{-1} \log Z(r,t)$ obeys
the evolution equation
\begin{equation}
\frac{\partial h}{\partial t} = \nu_0 \, \nabla^2
h  + \frac{\lambda_0}{2} \, ( \nabla h )^2 \, + V - \rho_0 \, \delta^{d'} (r)
\label{KPZ}
\end{equation}
with $\nu_0 = (2 \beta )^{-1}$ and $\lambda_0 =1$. This is the
Kardar-Parisi-Zhang (KPZ) equation of directed surface growth \cite{kpz.note}
with an additional term describing a local inhomogeneity in the rate of mass
deposition onto the surface.

Quite a few authors have studied these models. Early numerical work
for $d' =1$ indicates a delocalization transition at a finite defect strength
$\rho_0^c$ \cite{Kardar.unbinding,early.simulations}. In other large scale
simulations  \cite{TangLyuksyutov.unbind}, however,  it is found that
arbitrarily weak defects localize the polymer for $d'=1$, but a finite defect
strength is necessary for $d' >1$. This result is supported by an approximate
renormalization treatment for the problem on a hierarchical lattice
\cite{TangLyuksyutov.unbind},  by scaling arguments
\cite{BalentsKardar.letter},   and by an approximate functional renormalization
\cite{BalentsKardar.report}.  In \cite{KolomeiskyStraley.unbind} a Wilson type
renormalization is  discussed, but its consistency is unclear.  All of these
approaches rely on nonsystematic approximations; and the status of the
transition has remained controversial. The problem has so far defied attempts
at an exact solution even for $d' = 1$, in contrast to the related problem of
disorder-induced depinning from a rigid wall
\cite{Kardar.unbinding,KrugTang.unbinding}.

This letter is devoted to a field-theoretic study of the delocalization
transition. The zero-temperature fixed point of directed lines in a random
potential (with $\rho_0 = 0$) is a scale-invariant continuum field theory.  Its
two basic exponents,   the {\em roughness exponent} $\zeta$ and the  anomalous
dimension of the {\em disorder-averaged} free energy $-\omega$, are defined in
eqns. (\ref{diffcorr}) and (\ref{F}) below.  This theory is the starting point
for a systematic perturbation theory in the pinning potential; it involves an
$\varepsilon$-expansion with borderline dimension $d' =1$.  In contrast to
standard cases like $\phi^4$-theory,  here even the unperturbed system is a
field theory with complicated multipoint correlation functions, due to the
non-thermal averaging over  the disorder.  Nevertheless, two fundamental
properties of the local pinning field $\Phi(t)$ can be obtained in terms of the
exponents $\zeta$ and $\omega$: (i) its scaling dimension and (ii) the form of
its operator product expansion, see eqns. (\ref{x}) and (\ref{OPE}) below.
These properties determine the  renormalization group  equations for the
pinning strength to leading order, and hence the phase diagram of the system.
We find a transition at $\rho_0 = 0$ for $d' \leq 1$,  and at a finite
(nonuniversal) pinning strength $\rho_0^c$ for $d'=2$. The renormalization may
equivalently be carried out in the framework of the KPZ  dynamics, eq.
(\ref{KPZ}). This is discussed at the end of this letter, together  with some
implications for inhomogeneous growth processes.

To derive the renormalization group equations for the depinning problem, it is
necessary to study the polymer configurations in the unperturbed random system
(i.e. for $\rho_0 =0$).  The leading scaling behavior on large scales of
disorder-averaged correlation functions is due to the  sample-to-sample
fluctuations of the polymer paths of {\em  minimal energy}
\cite{FisherHuse.paths}.  Typical
transversal excursions of the paths, given e.g. by the two-point function
\begin{equation}
\Delta^2  (t_1 - t_2) \equiv
\overline{ \langle ( r(t_1) - r(t_2) )^2  \rangle }
\sim \vert t_1 - t_2  \vert^{2 \zeta} \, ,
\label{diffcorr}
\end{equation}
are characterized by the {\em roughness exponent} $\zeta$. It is larger than
for thermal fluctuations,
namely $\zeta = 2/3$ for $d' =1$ and $\zeta  \approx 5/8$  for $d' =2$ (see
e.g. \cite{KrugSpohn.review} and references therein).  The exponent $-\omega$
is
the anomalous dimension of the {\em  disorder-averaged} free  energy
$\overline{F} = - \beta^{-1}
\overline{ \log {\rm Tr} \exp (-\beta {\cal H}) }$,
whose universal part has the scaling form
\begin{equation}
\overline{F} (T,R) \sim \,
T^\omega \, {\cal F} (R / T^\zeta )
\label{F}
\end{equation}
in a finite system of transversal size $R$ and longitudinal size $T$.
Due to a ``tilt symmetry'', these exponents obey the relation
$\omega = 2 \, \zeta -1$
(see e.g. \cite{FisherHuse.paths,Schulz.RFIM,KrugSpohn.review}).

This asymptotic scaling is determined by a renormalization group fixed point
where the temperature $\beta^{-1}$ is an {\em irrelevant} scaling variable,
which nevertheless turns out to be crucial  to what follows. It governs the
crossover from the Gaussian fixed point   ($\sigma^2 = \rho_0 = 0$), that
describes the free thermal scaling on scales much smaller than the crossover
length $\tilde \xi_{\scriptscriptstyle \|}$ , to the  disorder fixed point. The
finite-size free energy (\ref{F}) has the crossover scaling form
$\overline{F} (T, \beta^{-1}) =
T^\omega \, \widetilde{{\cal F}} ( T / \tilde \xi_{\scriptscriptstyle \|} )$
at fixed $R / T^\zeta$. In the infrared, $\overline{F}$ is independent of
$\beta^{-1}$, in the ultraviolet, $\overline{F} \sim \beta^{-1} T^0 $.
Comparing
these two limits gives
$\beta^{-1}   \sim  \tilde {\xi_{\scriptscriptstyle \|}}^\omega$, i.e.
$-\omega$ is the anomalous dimension of   the temperature at the disorder fixed
point. In an analogous way, the two-point function (\ref{diffcorr}) has the
scaling form $\Delta^2 (t, \beta^{-1}) =
t^{2 \zeta} \,\,  {\cal J} ( \beta^{-1} t^{-\omega} ) $ with a
temperature-independent infrared limit and the ultraviolet asymptotic behavior
$ \Delta^2 ( t, \beta^{-1}) = \beta^{-1} t $ given by the Gaussian theory.
Hence the corrections to scaling of (\ref{diffcorr}) due to a small
temperature (which are illustrated in fig. 1)
are given by
\begin{equation}
\Delta^2 (t,\beta^{-1})  =
t^{2 \zeta} + \tilde{c} \, \beta^{-1} \,  t^{2 \zeta - \omega} + \dots
\label{correction.scaling}
\end{equation}
with $2 \zeta - \omega =  1$,
assuming analyticity of the scaling function ${\cal J}$.

Now we discuss the correlation functions of the local pinning field $\Phi (t)$
at the disorder fixed point.
The one-point function $\overline{ \left\langle \Phi (t)
\right\rangle }$  gives the probability that at time $t$, the polymer
is at the origin  $r=0$, averaged over thermal and disorder fluctuations.
In a system of infinite longitudinal size, but finite transversal size
$R$ (and periodic boundary conditions), one has by translational invariance
\begin{equation}
\overline{ \left\langle \Phi (t) \right\rangle } \, =
\, L^{-x}  \;\;\;\;  \mbox{with} \;\;\;\;  x = d' \, \zeta \, ,
\label{x}
\end{equation}
where $L$ is a longitudinal scale defined by $L \equiv R^{1/\zeta}$. In the
renormalization discussed below, $L$ has the r\^ole of an infrared cutoff,
and it generates the renormalization group flow.
The exponent $x$ is the {\em scaling dimension} of
the field  $\Phi$ at the disorder fixed point.

The {\em full} multipoint
correlation functions  $\overline{  \left\langle \Phi (t_1) \cdots \Phi (t_m)
\right\rangle }$  give the probability  that a (single) path crosses the line
$r=0$ at given times $t_1 \dots t_m$.  To discuss the short distance properties
of these objects, specifically  consider the two-point function  $\overline{
\left\langle \Phi (t_1) \Phi (t_2) \right\rangle }$ for  $\vert t_1 - t_2
\vert \ll L$. In this limit, it depends on the infrared cutoff as  $L^{-x}$,
i.e. in the same
way as the one-point-function  (\ref{x}). Hence asymptotically, it
factorizes into  $\overline{ \left\langle \Phi (t_1) \right\rangle }$ and the
$L$-independent ``return probability'' to the origin (which is  simply the
inverse spread of the paths $\sim \Delta (t)^{-d'}$ given by
(\ref{correction.scaling})),
\begin{equation}
\overline{ \langle \Phi (t_1) \Phi (t_2) \rangle }
\sim {\left\vert t_1 - t_2 \right\vert}^{-d' \zeta}
( 1 - \tilde{c} \beta^{-1} {\left\vert t_1 - t_2 \right\vert}^{-\omega}
  + \dots )
\overline{ \left\langle \Phi (t_1) \right\rangle } \, .
\label{SDA.full}
\end{equation}
Again the leading singularity
is due to the  sample-to-sample fluctuations of the minimal energy paths,
while the correction term is due to the thermal fluctuations around  these
paths. The leading, temperature-independent singularity equally occurs in the
{\em thermally disconnected} two-point function
$\overline{ \big\langle \Phi (t_1) \big\rangle \,  \big\langle \Phi
(t_2) \big\rangle } $ \cite{singularities.note}.
Hence in the {\em connected} two-point function
$ \overline{ \left\langle \Phi (t_1) \, \Phi (t_2) \right\rangle^c } $, only
the subleading singularity survives. An analogous argument applies to the
singularities in any correlation function
$ \overline{ \langle \dots \Phi(t) \Phi (t') \dots
             \rangle } $ as $ | t - t' | \to 0 $. Therefore the relation
\begin{equation}
\Phi (t) \Phi (t') \sim  c \,  \beta^{ \, -1 }   {
\left\vert t - t' \right\vert }^ {- x -\omega} \, \Phi (t)
\label{OPE}
\end{equation}
(with a constant $c>0$) is valid as an operator identity, i.e. inserted into
an arbitrary  {\em connected} correlation function
$\overline{ \langle \dots \Phi(t) \Phi (t') \dots \rangle^c } $. The notion of
an {\em operator algebra} that encodes the universal short-distance properties
of correlation functions is familiar in field theory \cite{ope}. The new
feature of (\ref{OPE}) is that the leading singularity is governed by a
correction-to-scaling exponent. This is a consequence of the breakdown of
hyperscaling at the zero-temperature disorder fixed point and has direct
implications for the renormalization of the pinning problem to which we now
turn.

The renormalized perturbation theory for the Hamiltonian (\ref{H}) is
constructed along the lines
of ref. \cite{Laessig.roughening}. The universal part of the disorder-averaged
free energy density
$\overline{f} \equiv \lim_{T \to \infty} \overline{F}\, / T   R^{d'} $
can be expanded in powers of the dimensionful bare coupling constant $\rho_0$,
\begin{equation}
\overline{f (\rho_0,L)} -  \overline{f(0,L)} =
\beta^{-1}  R^{-d'}  \sum_{m = 1}^\infty
\frac{(-\beta \rho_0)^m}{m !} \, J_m \, ,
\label{expansion}
\end{equation}
where
\begin{equation}
J_m = \int {\rm d} t_2 \cdots {\rm d} t_m
\overline{ \left\langle  \Phi (0)  \Phi (t_2) \cdots \Phi (t_m)
\right\rangle^c  }   \, .
\end{equation}
A weak defect potential distorts the minimal energy paths of the
unperturbed  system, the dominant paths reorganize exploiting the low-lying
excitations.  As discussed above, the  statistics of these excitations is
encoded in  the {\em connected} correlation functions that appear in $J_m$.
The leading ultraviolet singularities of these integrals are dictated by
the operator algebra (\ref{OPE}). In an analytic continuation
to arbitrary $d'$,  these show up as poles in
$\varepsilon ( d') \equiv 1 - (d' \zeta + \omega)$, which serves as expansion
parameter. Inserting (\ref{x}) and (\ref{OPE}) into
(\ref{expansion}), we find to second order
\begin{equation}
\overline{f(\rho_0,L) } - \overline{f(0,L)}  =
- L^{\omega - 1} \overline{ \left\langle \Phi \right\rangle }
\left ( w_0 - \frac{c}{ \varepsilon ( d' ) }  {w_0}^2 \right )
 +  O \left( {w_0}^3 , \varepsilon^0 {w_0}^2 \right)  .
\label{ren}
\end{equation}
where $w_0 \equiv \rho_0 L^{\varepsilon}$ is the dimensionless bare coupling
constant. The pole in $\varepsilon ( d')$ can be absorbed into a renormalized
coupling constant $w = Z(w) w_0$ with
$Z(w) = 1 - (c / \varepsilon)  w + O \left( w^2 \right) $. Its renormalization
group flow \cite{higher_orders}
\begin{equation}
 L \partial_L w =
\varepsilon (d')   w - c  w^2  + O \left( w^3 \right)
\label{RGE}
\end{equation}
determines the large scale behavior of the perturbed system.
For $\varepsilon (d') > 0$, i.e. for  $d' <1$, the perturbation is relevant:
for any attractive bare defect potential, the
renormalized coupling is driven towards large attractive values. Hence the
flux line is localized by an arbitrary weak attractive columnar defect.
The localization length diverges as
$\xi_\perp \sim \vert \rho_0 \vert^{- \nu_\perp}$ with
$\nu_\perp = \zeta (d')  / \varepsilon (d')$ when the defect strength
approaches zero from below.
In the borderline dimension  $d' = 1$ an attractive  defect potential is
marginally relevant: the line is still localized by an
arbitrary weak columnar defect, but with an essential singularity in the
localization length $\xi_\perp \sim \exp ( 2 / 3 c \vert w \vert)$.

For $\varepsilon (d') < 0$, i.e. for $d' >1$, a {\em weak} defect is an
irrelevant perturbation. The transition to a pinned state now takes place
at a {\em finite} critical strength $\rho_0^c$ (which however depends on the
microscopic scales of the system and is hence nonuniversal). It is governed
by the nontrivial fixed point $w^\ast = \varepsilon (d') / c < 0$ of
(\ref{RGE}). Close to the transition, the localization length diverges as
$\xi_\perp \sim | \rho_0  - \rho_0^c |^{- \nu_\perp}$,
where $\nu_\perp = \zeta (d') / y^\ast (d') $ and $y^\ast (d')$
is given by the $\varepsilon$-expansion
$y^\ast =  - \varepsilon (d')  + O(\varepsilon^2)$.

Additional insight into this problem may be gained by the mapping onto the  KPZ
equation (\ref{KPZ}). From this stochastic equation, one constructs in a
standard way  the generating functional
$ {\rm Tr} \exp( - S[h,\widetilde h]) $
of the dynamical correlation functions
(denoted by $ \langle \! \langle \dots \rangle \! \rangle  $)
in terms of the height field $h$ and the ``conjugate'' field $\widetilde h$
\cite{generating.functional}. Insertions of this field generate response
functions, e.g.
$  \langle \! \langle h(r,t) \, \Pi_i \widetilde{h} (r_i, t_i)
   \rangle \! \rangle =
   \langle \! \langle \delta \, h(r,t) / \Pi_i \delta \, V(r_i,t_i)
   \rangle \! \rangle $.
In the dynamical action, there is a  term
$ S_i = \rho_0 \int {\rm d} t \widetilde h (0,t) $,
the analogue of the pinning term in (\ref{H}). The perturbation series
(\ref{expansion}) for the restricted free energy
$ \beta^{-1} \overline{ \log Z(r,t)} =
  \langle \! \langle h(r,t)  \rangle \! \rangle $
is in one-to-one correspondence with the dynamical perturbation series;
we have
\begin{equation}
\beta^{m-1} \;\;
\overline{ \langle \Phi (t_1) \cdots \Phi (t_m ) \rangle^c } =
\langle \! \langle h(r,t) \widetilde{h} (0, t_1) \cdots \widetilde{h} (0, t_m)
\rangle \! \rangle \, .
\label{moments}                                                             
\end{equation}
Hence $ \widetilde h(0,t) = \beta \Phi (t) $ is a field of scaling dimension
$\widetilde{x} = d' \, \zeta + \omega $; by virtue of (\ref{OPE}), it obeys
the short-distance algebra
\begin{equation}
\widetilde{h} (0, t) \widetilde{h} (0, t') \sim
c\, \vert t-t' \vert^{-\widetilde{x}} \; \widetilde{h} (0, t') + \dots \; .
\label{dynOPE}
\end{equation}
Its leading singularity is no longer a correction-to-scaling
exponent; the peculiarity of the correlation functions written in terms of the
$\widetilde{h} $ fields is rather that they have to be computed in the
nontrivial
``vacuum'' state $h(r,t)$. This makes the {\em nonunitarity} of this theory
manifest, which is generated by the averaging over disorder.

For $\rho_0 < 0$, the term $S_i$ describes an excess mass deposition onto the
growing surface at $ r =  0 $. This term breaks the translational invariance.
For stationary growth
$ \langle \! \langle h(r,t) \rangle \! \rangle = v t + H(r) $ in a
one-dimensional system of size $R$, it results in an excess growth velocity
and an approximately triangular surface profile $H(r)$
\cite{inhomogeneous.KPZ}.
{}From the mapping to the polymer system, one concludes that for
$\xi_\perp (w) \ll R$, the excess velocity scales as
$ (v(w,R) - v(0,R)) \sim \xi_{\perp}^{(-1 + \omega)/ \zeta}
  \sim \exp (2 / 3 c w)$.
The same essential singularity shows up in the slope
$ \vert \partial H / \partial r \vert \sim (v(w,R) - v(0,R))^{1/2} $.
The response function has the form
$ \langle \! \langle h(r, t) \widetilde{h} (r_1,t_1) \rangle \! \rangle =
\xi_\perp^{-1} {\cal G} (r_1 / \xi_\perp)$
for $t_0 \to \infty$ and $ r, \xi_{\perp} \ll R$.
All of these quantities are accessible in numerical simulations which could
provide a useful test of the results discussed in this letter.

In summary, we have shown that a class of field theories with quenched
randomness shares with conventional field theories the notion of a
short-distance algebra of its scaling operators, whence they are amenable
to systematic renormalization. \\
\\
{\em Note added:}
In the final stages of this work we received a preprint by Hwa and
Nattermann \cite{HwaNattermann} on the depinning problem.
They exploit the mapping onto the KPZ equation (\ref{KPZ}).
In $d'=1$, they obtain the algebra (\ref{dynOPE}) by using a mode-coupling
approximation for the dynamical correlation functions. \\
\\
We thank J. Krug, R. Lipowsky, and L.-H. Tang for useful discussions.

\newpage

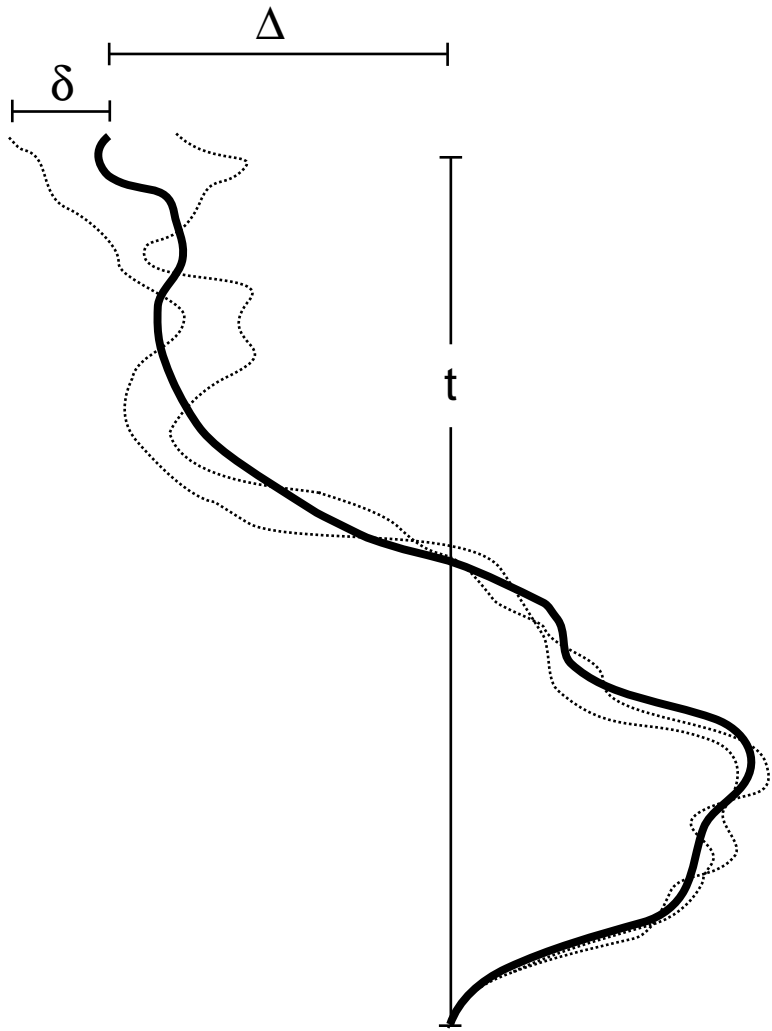
\noindent {\bf Figure 1}

\noindent Ensemble of paths with a common starting point in a typical
disordered sample.
The thick line denotes the path of minimal energy; its transversal
fluctuations are given by $\Delta^2 (t) \sim t^{2 \zeta} $.
The thin lines represent thermal fluctuations of width
$\delta^2 (t) \equiv \overline{ \langle r^2 (t) \rangle^c } \sim \beta^{-1} t$
around the minimal path.


\newpage


\begin{thebibliography}{99}
\baselineskip=10pt
\footnotesize{


\bibitem{Interfaces.review}
G. Forgacs, R. Lipowsky, and T.M. Nieuwenhuizen,  in {\em Phase transitions and
Critical Phenomena}, Vol. 14,  ed. C. Domb and J. Lebowitz (Academic Press,
London, 1991).


\bibitem{HuseHenley.roughening}
D.A. Huse and C.L. Henley, Phys. Rev. Lett. 54 (1985), 2708.


\bibitem{fluxlines}
M.V. Feigel'man {\em et al.}, Phys. Rev. Lett. 63 (1989), 2303;
J.P. Bouchaud, M. M\'ezard, and J.S. Yedidia, Phys. Rev. Lett. 67 (1991), 3840;
Phys. Rev. B 46 (1992), 14686.


\bibitem{KPZ}
M. Kardar, G. Parisi, and Y.-C. Zhang, Phys. Rev. Lett.  56 (1986), 889.


\bibitem{KrugSpohn.review}
J. Krug and H. Spohn, in {\em Solids Far From Equilibrium: Growth, Morphology
and Defects}, ed. C. Godr\`eche (Cambridge, 1990).


\bibitem{noisy.Burgers}
D.A. Huse, C.L. Henley, and D.S. Fisher, Phys. Rev. Lett. 55 (1985), 2924;
D. Forster, D.R. Nelson, M. Stephen, Phys. Rev. A 16 (1977), 732.


\bibitem{relation.spinglass}
B. Derrida and H. Spohn, J. Stat. Phys. 51 (1988), 817;
M. M\'ezard and G. Parisi, J. Phys. I (France) 1 (1991), 809;
H. Kinzelbach and H. Horner, J. Phys. I (France) 3 (1993), 1329 and 1901.


\bibitem{experiments.flux}
L. Civale {\em et al.}, Phys. Rev. Lett. 67 (1991), 648;
R.C. Budhani, M. Suenaga, and S.H. Liou, Phys. Rev. Lett. 69 (1992), 3816;
S. Behler {\em et al.}, Phys. Rev. Lett. 72 (1994), 1750.


\bibitem{inhomogeneous.KPZ}
D.E. Wolf and L.-H. Tang, Phys. Rev. Lett. 65 (1990), 1591;
S.A. Janowsky and J.L. Lebowitz, Phys. Rev. A 45 (1992), 618.


\bibitem{kpz.note}
The two basic exponents of the KPZ universality class for $\rho_0 = 0$,
the roughness exponent $\chi$ and the dynamical exponent $z$,
are defined by
$ \langle \! \langle (h(0,0) - h(r, t))^2 \rangle \! \rangle \sim
|r|^{2\chi} f(|t r^{-z}|)$,
and are related to the polymer exponents by
$\chi = \omega / \zeta$ and
$ z = 1 / \zeta$.


\bibitem{Kardar.unbinding}
M. Kardar, Phys. Rev. Lett. 55 (1985), 2235;
Nucl. Phys. B290 (1987), 582.


\bibitem{early.simulations}
J. Wuttke and R. Lipowsky, Phys. Rev. B 44 (1991), 13042;
M. Zapotocky and T. Halpin-Healy, Phys. Rev. Lett. 67 (1991), 3463.


\bibitem{TangLyuksyutov.unbind}
L.-H. Tang and I.F. Lyuksyutov, Phys. Rev. Lett. 71 (1993), 2745.


\bibitem{BalentsKardar.letter}
L. Balents and M. Kardar, Europhys. Lett. 23 (1993), 503 generalize a
scaling argument by
T. Nattermann and R. Lipowsky, Phys. Rev. Lett. 61 (1988), 2508.


\bibitem{BalentsKardar.report}
L. Balents and M. Kardar, MIT report (1993), to be published.


\bibitem{KolomeiskyStraley.unbind}
E.B. Kolomeisky and J.P. Straley, Univ. of Kentucky report, 1992;
E.B. Kolomeisky and J.P. Straley, Univ. of Kentucky report, 1994.


\bibitem{KrugTang.unbinding}
J. Krug and L.-H. Tang, Phys. Rev. E, in press.


\bibitem{FisherHuse.paths}
D.S. Fisher and D.A. Huse, Phys. Rev. B 43 (1991), 10728.


\bibitem{Schulz.RFIM}
U. Schulz {\em et al.}, J. Stat. Phys. 51 (1988), 1.


\bibitem{singularities.note}
In general there are also contributions of paths crossing $r=0$ at one of
the two times, but the leading singularity for $\vert t_1 - t_2 \vert \to 0$
again comes from  paths that cross the line at both times.


\bibitem{ope}
For example, the algebra of the pinning field at the Gaussian fixed point,
which is relevant to temperature-driven unbinding transitions, reads
$  \Phi ( t ) \Phi (t') \sim |t - t'|^{-x_0} \Phi (t) + \dots $
with $ x_0 = d'/2$. See ref. \cite{Laessig.roughening}.


\bibitem{Laessig.roughening}
M. L\"assig and R. Lipowsky, Phys. Rev. Lett. 70 (1993), 1131;
M. L\"assig and R. Lipowsky, in {\em Fundamental Problems in Statistical
Mechanics VIII}, ed. H. van Beijeren, North Holland, Amsterdam, 1994.


\bibitem{higher_orders}
Further primitive singularities are expected at higher orders in the
perturbation expansion; hence in contrast to thermal depinning
\cite{FERMIONS}, the flow equation does not terminate at this order.


\bibitem{FERMIONS}
M. L\"assig, J\"ulich report (1993).


\bibitem{generating.functional}
For the general formalism, see e.g.
P.C. Martin, E.D. Siggia, H.A. Rose, Phys. Rev. A8 (1973) 423;
R. Bausch, H.K. Janssen, H. Wagner, Z. Phys. B24 (1976), 113.


\bibitem{HwaNattermann}
T. Hwa and T. Nattermann, Princeton report, cond-mat 9404031 (1994).


}

\end{thebibliography}
\end{document}